\documentclass[graybox]{svmult}

\usepackage{mathptmx}       
\usepackage{helvet}         
\usepackage{courier}        
\usepackage{type1cm}        
\usepackage{makeidx}         
\usepackage{graphicx}        
\usepackage{multicol}        
\usepackage[bottom]{footmisc}

\makeindex             

\begin{document}

\title*{Symbolic Toolkit for Chaos Explorations}
\author{Tingli Xing, Jeremy Wojcik, Roberto Barrio and Andrey Shilnikov}
\institute{Tingli Xing, Jeremy Wojcik and Andrey Shilnikov  
\at Neuroscience Institute and Department of  Mathematics
and Statistics, Georgia State University, Atlanta 30303, USA, \email{ashilnikov@gsu.edu}
\and Roberto Barrio \at Departamento de Matem\'atica Aplicada and IUMA,
University of Zaragoza. E-50009. Spain}
%
%
\maketitle

\abstract*{New computational technique \cite{BSS2012b,BSS2012a} based on the symbolic description utilizing kneading invariants is used for the exploration of parametric chaos in two exemplary systems with the Lorenz attractor: a normal model from mathematics, and a laser model from nonlinear optics. The technique uncovers the stunning complexity and universality of the patterns discovered  in the bi-parametric scans of the given models and detects their organizing centers -- codimension-two T-points and separating saddles.}

\abstract{New computational technique  \cite{BSS2012b,BSS2012a}  based on the symbolic description utilizing kneading invariants is used for explorations of parametric chaos in a two exemplary systems with the Lorenz attractor: a normal model from mathematics, and a laser model from nonlinear optics. The technique allows for uncovering the stunning complexity and universality of the patterns discovered in the bi-parametric scans of the given models and detects their organizing centers -- codimension-two T-points and separating saddles.}

\section{Introduction}
\label{sec:1}

Several analytic and experimental studies, including modeling simulations, have focused
on the identification of  key signatures to serve as structural invariants. Invariants would allow dynamically similar nonlinear systems with chaotic dynamics from diverse origins to be united into a single class. Among these key structures are various homoclinic and heteroclinic bifurcations of low codimensions that are the heart of the understanding of complex behaviors because of their roles as organizing centers of dynamics in parameterized dynamical systems.

\begin{figure}[hbt!]
 \begin{center}
 \includegraphics[width=.6\textwidth]{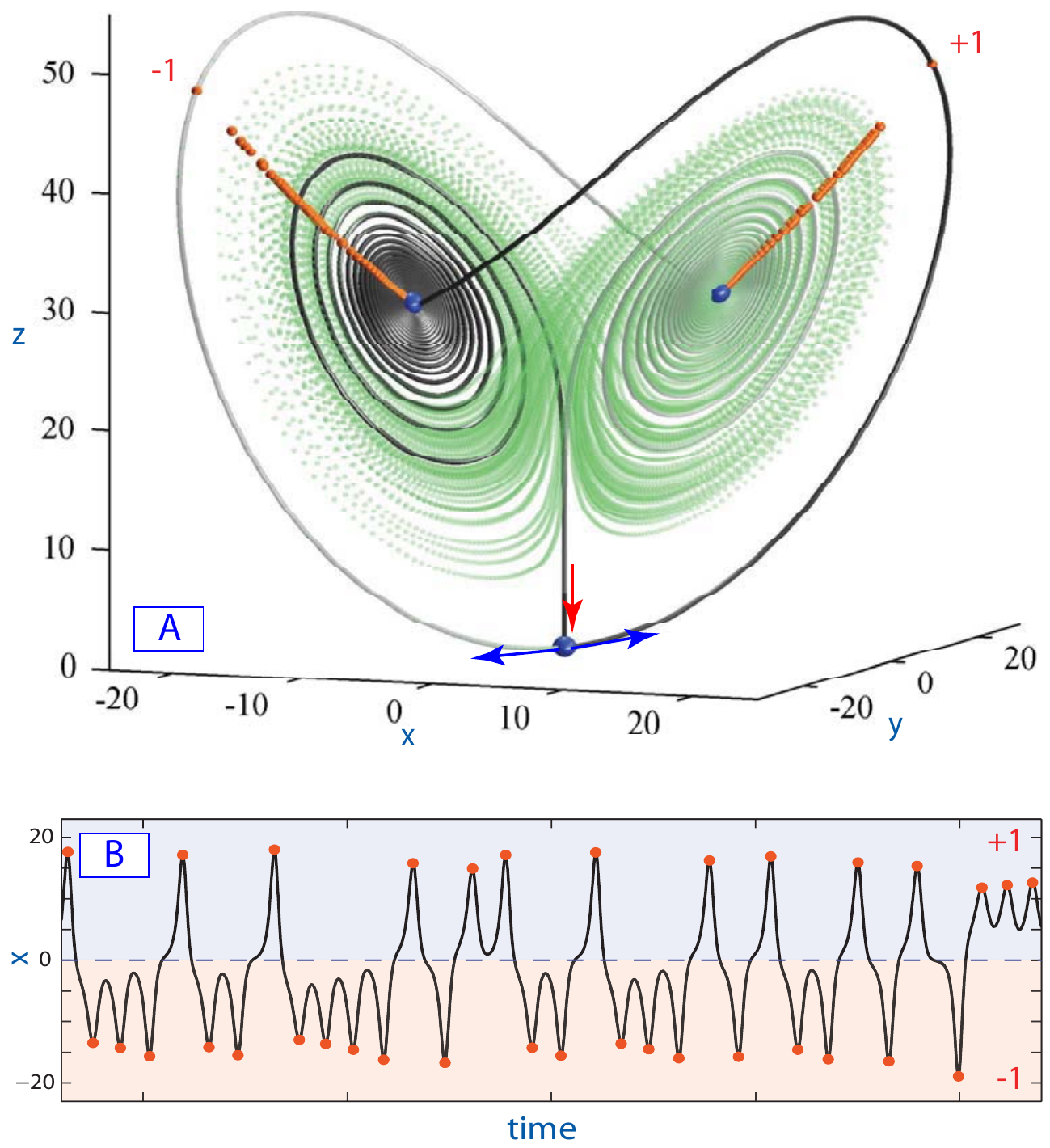}
\caption{(A) Heteroclinic connection (in dark color) between the saddle at the origin and two saddle-foci (blue spheres) being overlaid with the strange attractor (green light color) on the background at the primary T-point $(r=30.38,\,\sigma=10.2)$ in the Lorenz model. Orange spheres on the butterfly wings indicating the turning points around the right and left saddle-foci define the kneading sequence entries, $\{\pm 1\}$, respectively. (B) A typical time evolution of either symmetric coordinate of the right separatrix of the saddle.}\label{fig1l}
 \end{center}ƒ
\end{figure}

One computationally justified approach for studying complex dynamics capitalizes on the sensitivity of deterministic chaos. Sensitivity of chaotic trajectories can be quantified in terms of the divergence rate evaluated through the largest Lyapunov characteristic exponent. In several low-order dissipative systems, like the R\"ossler model, the computational technique based on  the largest Lyapunov characteristic exponent reveals that they possess common, easily recognizable patterns involving spiral structures in bi-parametric planes \cite{BYK93,SST93}. Such patterns have turned out to be ubiquitously in various discrete and continuous-time  systems \cite{Lor08,GN84,BBS09a}, and they are easily located, as spiral patterns have regular and chaotic spiral ``arms'' in the systems with the Shilnikov saddle-focus \cite{sh65,BBSS11,BSS2012b,BSS2012a}.

Application of the Lyapunov exponents technique fails, in general, to reveal fine structures embedded in the bi-parametric scans of Lorenz-like systems.  This implies that the instability of the Lorenz attractors does not vary noticeably as control parameters of the system are varied. This holds true when one attempts to find the presence  of characteristic spiral structures that are known to theoretically exist in Lorenz-like systems \cite{BYK93,GS86}, identified using accurate bifurcation continuation approaches \cite{ALS91,SST93}. Such spirals in a bi-parametric parameter plane of a Lorenz-like system are organized around the T[erminal]-points; corresponding to codimension-two, closed heteroclinic connections involving two saddle-foci and a saddle at the origin, see Fig.~\ref{fig1l}. Such T-points have been located in various models of diverse origins including electronic oscillators \cite{Bykov98,FFR02} and nonlinear optics \cite{FMH91}.

Despite the overwhelming number of studies reporting the occurrence of various spiral structures, there is yet 
little known about construction details and generality of underlying bifurcation scenarios which gives 
rise to such spiral patterns. Additionally, little is known how such  patterns are embedded in the parameter space of the models with 
the Lorenz attractors. Here we present a computational toolkit capitalizing on the symbolic representation for the dynamics of Lorenz-like systems that employ kneading invariants \cite{MT88}. We will then show how the toolkit detects various fractal structures in bi-parametric scans of two exemplary systems: a normal model from mathematics, and a laser model from nonlinear optics. For the further details we refer the reader to the original paper \cite{BSS2012a}.

\section{Kneading invariants for a Lorenz like system}
\label{sec:2}

Chaos can be quantified by several means. One customary way is through the evaluation of topological entropy.
The greater the value of topological entropy, the more developed and unpredictable the chaotic dynamics become. Another practical approach for measuring chaos in simulations capitalizes on evaluations of the largest (positive) Lyapunov exponent of a long yet finite-time transient on the chaotic attractor.

A trademark of any Lorenz-like system is the strange attractor of the iconic butterfly shape, such as shown in 
Fig.~\ref{fig1l}. The ``wings'' of the butterfly are marked with two symmetric ``eyes'' containing equilibrium states, stable or not, isolated from  the trajectories of the Lorenz attractor. This attractor is structurally unstable \cite{GW79,ABS83} as it bifurcates constantly as the parameters are varied. The primary cause of structural and dynamic instability of chaos in the Lorenz equations and similar models is the singularity at the origin -- a saddle with two one-dimensional outgoing separatrices. Both separatrices densely fill the two spatially symmetric wings of the Lorenz attractor in the phase space. The Lorenz attractor  undergoes a homoclinic bifurcation when the separatrices of the saddle change the alternating pattern of switching between the  butterfly wings centered around the saddle-foci. At such a change, the separatrices comes back to the saddle thereby causing a homoclinic explosions in phase space \cite{ABS77,KY79}.

\begin{figure*}[!htb]
\begin{center}
\includegraphics[width=0.99\textwidth]{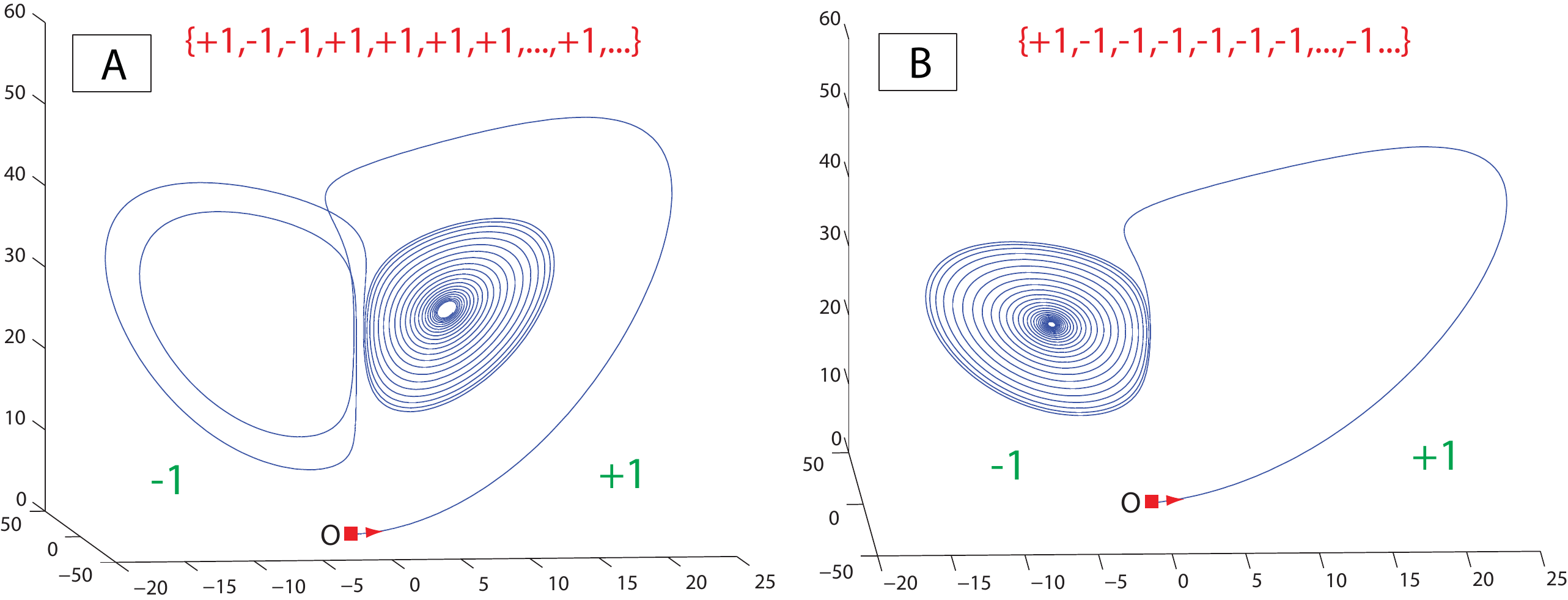}
\caption{\label{figtwo}  Truncated kneading sequences generated by the right outgoing separatrix of the saddle at the origin in a typical Lorenz-like equation at two distinct values of the parameters.}
\end{center}
\end{figure*}

The time progression of the ``right” (or symmetrical ``left”) separatrix  of the origin can be described geometrically and categorized in terms of the number of alternations around the nonzero equilibrium states  in the phase space of the Lorenz-like system (Fig.~\ref{fig1l}). Alternatively, the description can be reduced to  the time-evolution of a coordinate of the separatrix,  as shown in panel~B of Fig.~\ref{fig1l}. The sign-alternation of the $x$-coordinate suggests the introduction of a $\{\pm 1\}$-based alphabet  for the symbolic description of the separatrix. Namely, whenever the right separatrix turns around $O_1$ or $O_2$, we record $+1$ or $-1$, respectively. For example, the time series shown in panel~B generates the following kneading sequence starting with
$\{ +1, -1,-1,-1,+1,-1,-1,+1,-1, \ldots \} $.

We introduce and demonstrate a new computational toolkit for the analysis of chaos in the Lorenz-like models. The toolkit 
is inspired by the idea of kneading invariants introduced in \cite{MT88}. A kneading invariant is a quantity that is intended to uniquely describe
 the complex dynamics of the system that admit a symbolic description using two symbols, here $+1$ and $-1$. 
 
The  kneading invariant for either separatrix of the saddle equilibrium state  of the Lorenz attractor  can be defined in the form of a formal power series:
\begin{equation}\label{fs}
P(q) = \sum_{n=0}^\infty {\kappa}_{n}\,q^n .
\end{equation}
Letting $q\in (0,\,1)$ guarantees the series is convergent. The smallest zero, $q^*$, if any, of the graph of (\ref{fs}) in the interval $q\in (0,\,1)$ yields the topological entropy, $h(T) = \ln(1/q^*)$. 

The kneading sequence $\{{\kappa}_{n}\}$ composed of only $+1$s  corresponds to the ``right" separatrix of the saddle converging to an $\omega$-limit set with $x(t)>0$, such as a stable focus or stable periodic orbit.
The corresponding kneading invariant is maximized at $\{P_{\rm max}(q)\}=1/(1-q)$. When the right separatrix  converges to an attractor with $x(t)<0$, then the kneading invariant is given by $\{P_{\rm min}(q)\}=1-q/(1-q)$ because the first entry $+1$ in the kneading sequence is followed by infinite $-1$s. Thus, $\left [\{P_{\rm min}(q)\},\,\{P_{\rm max}(q)\} \right ]$ yield the range of the kneading invariant values; for instance $\left [\{P_{\rm min}(1/2)\}=0,\,\{P_{\rm max}(1/2)\}=2 \right ]$.

\begin{figure*}[ht!]
 \begin{center}
\includegraphics[width=.4\textwidth]{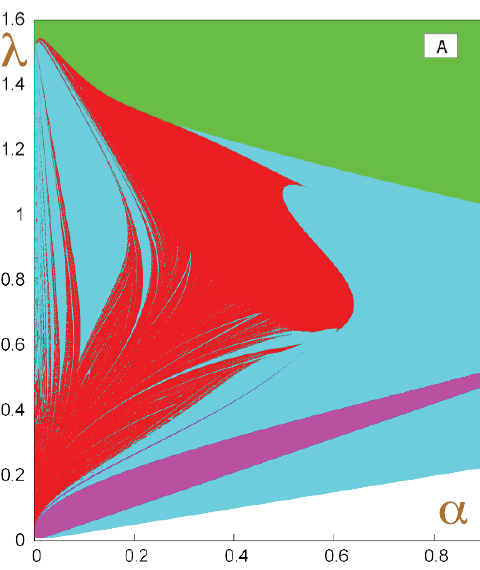}~~\includegraphics[width=.6\textwidth]{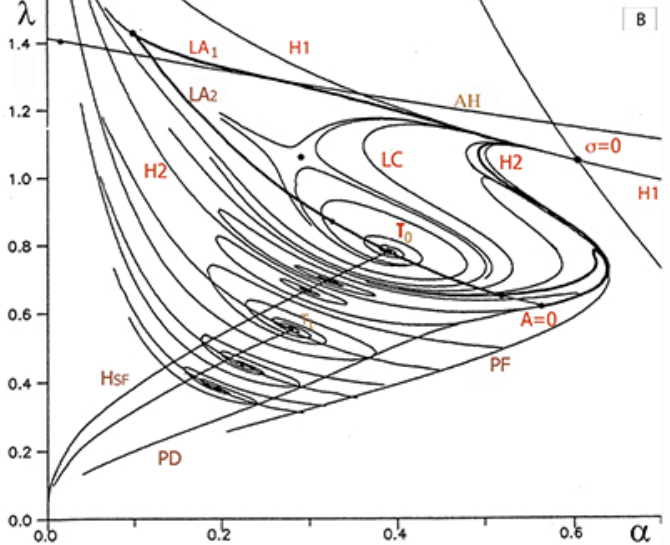}
\caption{\label{fig7l} (left) Existence regions of attractors of the Shimizu-Morioka model the in $(\alpha,\,\lambda)$ being differentiated by the sign of the largest Lyapunov exponent, $L_{max}$. Color legend for the attractors of the model: green--stable equilibrium states, $L_{max}<0$; blue --
stable periodic orbits with a nodal normal behavior, $L_{max}=0$; magenta -- a periodic orbit with a focal normal behavior; red -- chaotic attractors with $L_{max}>0$, with identified lacunae.
Courtesy of \cite{GOST05}. (right)  Detailed $(\alpha,\lambda)$-parameter plane of the Shimizu-Morioka model obtained by the parameter continuation method (courtesy of \cite{SST93}). Legend: $AH$ stands for a supercritical Andronov-Hopf
bifurcation, $H_1$ stands for the homoclinic butterfly made of two separatrix loops; the codimension-two points corresponding to the resonant saddle $\sigma=0$ on $H_1$ organizes the bifurcation unfolding of the model; cod-2 point $A=0$ stands for an orbit-flip bifurcation for the double-loop homoclinics on $H_2$. The thick line demarcates, with good precision, the existence region of the Lorenz attractor bounded by $LA_1$ and $LA_2$.}
 \label{simo}
 \end{center}
 \end{figure*}

Two samples of the separatrix pathways shown in Fig. \ref{figtwo} 
generating the following kneading invariants illustrate the idea.
$$
\begin{array}{l}  P_{A}(1/2)=+1-1/2-1/4+1/8+1/16+1/32+1/64 \ldots +1/2^n \ldots = 1/2,\\
P_B(1/2)= +1-1/2-1/4-1/8-1/16-1/32-1/64  \ldots -1/2^n \ldots = 0,
\end{array}
$$

In computational studies of the models below, we will consider a partial kneading power series truncated to the first
20 entries: $P_{20}(q) = \sum_{n=0}^{20} {\kappa}_{n}\,q^n $.
The choice of the number of entries is not motivated by numerical precision, but by simplicity, as well as by resolution of the bitmap mappings for the bi-parametric scans of the models. One has also to determine the proper value of $q$: setting it too small makes the convergence
 fast so that the tail of the series has a little significance and hence does not differentiate the fine dynamics of the Lorenz equation for longer kneading sequences.

At the first stage of the routine, we perform a bi-parametric scan of the model within a specific range in the parameter plane. The resolution of scans is set by using mesh grids of $[1000 \times 1000]$ equally distanced points. Next by integrating the same separatrix of the saddle point we identify and record the sequences $\{\kappa_n\}_{20}$ for each point of the grid in the parameter plane. Then we define the bi-parametric mapping: for the Shimizu-Morioka model below it is  $(\alpha,\lambda) \to  P_{20} (q)$ for some chosen  $q$, the value of which determines the depth of the scan.
The mapping is then colorized in Matlab by using various built-in functions ranging between to $P_{20}^{\rm min}$ and  $P_{20}^{\rm max}$, respectively.  In the mapping, a particular color in the spectrum  is associated with a persistent value of the kneading invariant on a level curve. Such level curves densely foliate the bi-parametric scans.


\section{Kneading scans of the Shimizu-Morioka model}
\label{sec:3}

Here we will examine the kneading-based bi-parametric scanning of the Shimizu-Morioka model \cite{SM80,ALS91}:
\begin{equation}
\dot{x} = y, \quad \dot{y} = x -\lambda y -xz, \quad \dot{z} = - \alpha z+ x^2; \label{sm}
\end{equation}
with $\alpha$ and $\beta$ being positive bifurcation parameters. The $Z_2$-symmetric model has three equilibrium states: a simple saddle, with one-dimensional separatrices, at the origin, and two symmetric stable-foci which can become saddle-foci through a supercritical Andronov-Hopf bifurcation.

This model was originally introduced to examine a pitch-fork bifurcation of the stable figure-8 periodic orbit that gives rise to multiple cascades of period doubling bifurcations in the Lorenz equation at large values of the Reynolds number. It was proved in \cite{SST93} that the Eqs.~(\ref{sm}) would be a universal normal form for several codimension-three bifurcations of equilibria and periodic  orbits on $Z_2$-central  manifolds.
The model turned out to be very rich dynamically: it exhibits various interesting global bifurcations \cite{ASHIL93} including T-points for heteroclinic connections.

In the case study of the Shimizu-Morioka model, we compare the proposed kneading scanning apparatus with the customary bi-parametric sweeping based on the evaluation of the Lyapunov exponent spectrum computed over a finite time interval. The latter is shown in Fig.~\ref{simo} \cite{GOST05}. 

\begin{figure}[ht!]
 \begin{center}
\includegraphics[width=1.0\textwidth]{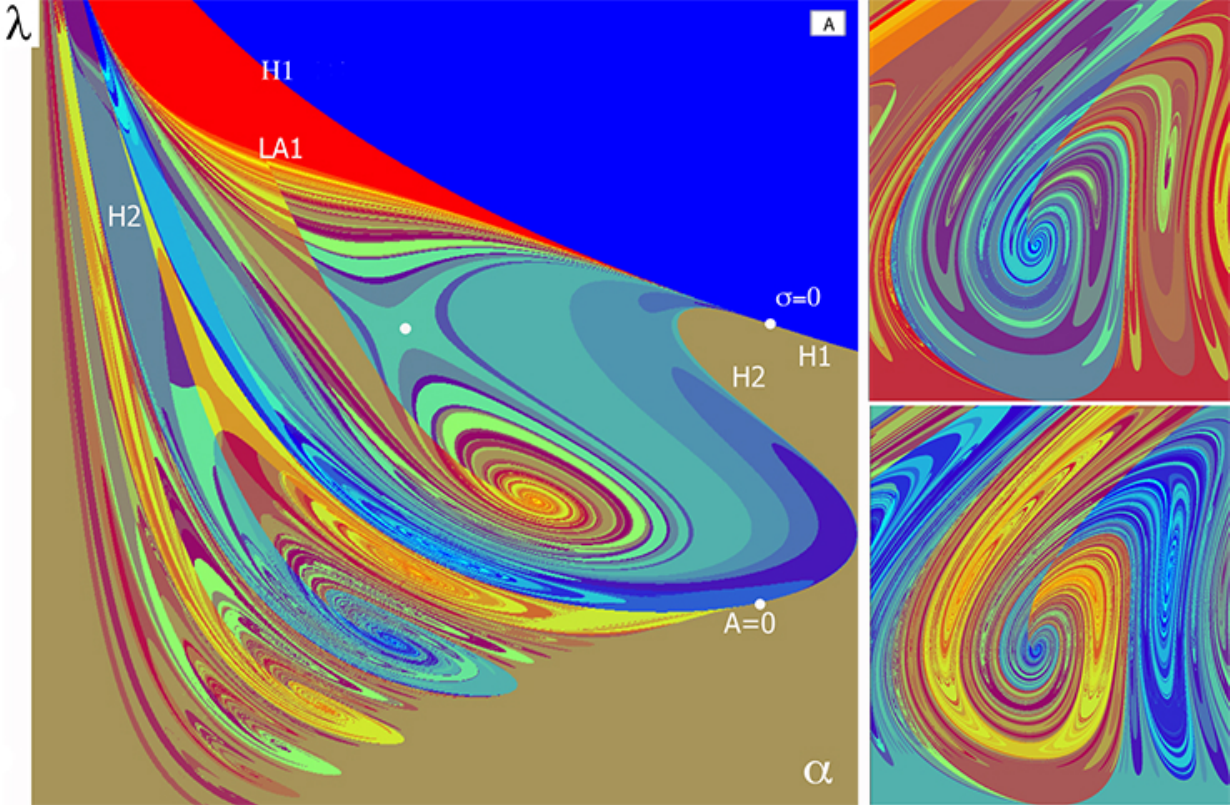}
\caption{\label{fig5l} (left) The scan revealing multiple T-points and saddles that  globally organize complex chaotic dynamics of the Shimizu-Morioka model. Solid-color regions associated with constant values of the kneading invariant correspond to simple dynamics dominated by stable equilibria (brown) or stable periodic orbits (blue). The border between the brown and blue regions corresponds to the bifurcation curve of the homoclinic butterfly. The codimension-two
point, $\sigma=0$, gives rise to loci of bifurcation curves including $LA_1$ below which the Lorenz attractor exists. Bifurcation loci of the other codimension-two point, $A=0$ (yellow zone)  giving rise to subsidiary orbit-flip bifurcations on turns of spirals around T-points, are separated by saddles (two large scale ones) in the parameter plane. (right) Two zoom of distinct scanning depths of the $(\alpha,\lambda)$-parametric mapping near the secondary T-point, $T_{1}$,  revealing fine fractal structures embedding smaller-scale spirals around several visible descendant T-points.}
 \end{center}
\end{figure}

The regions of the solid colors are associated with the sign of the largest Lyapunov exponent, $L_{\rm max}$: $L_{\rm max} < 0$ values correspond
to steady state attractors in the green region; $L_{\rm max}=0$ corresponds to periodic attractors in the blue region; and $L_{\rm max}>0$  is associated with chaotic dynamics in the model in the region. Note the blue islands in the red-colored region that correspond to stability windows in chaos-land. Such windows in the Lorenz attractor have an emergent lacuna containing one more three periodic orbit. Other than lacunas, the diagram shows no sign of any structure in the red region corresponding to the chaotic dynamics.

Indeed, the structure of the bifurcation set of the Shimizu-Morioka is very complex. The detailed bifurcation diagram is shown in the top panel of Fig.~\ref{fig5l}. It reveals several T-points,  and multiples  curves corresponding to an Andronov-Hopf (AH), pitch-fork (PF), period doubling (PD) and homoclinic (H) bifurcations that shape the existence region of the Lorenz attractor in the model.   
The detailed description of the bifurcation structure of the Shimizu-Morioka model is out of scope of this paper. The reader can find a wealth of information on bifurcations of the Lorenz attractor in the original papers \cite{ASHIL93,SST93}. We point out that those bifurcation curves were continued in the $(\alpha,\lambda)$-parameter plane by following various bifurcating solutions, as periodic, homo- and heteroclinic orbits, in the phase space of the model.

The panel~A of Fig.~\ref{fig5l} is a {\em de-facto} proof of the new kneading invariant mapping technique. The panel represents the color bi-parametric scan of the dynamics of the Shimizu-Morioka model that is based on the evaluation of the first 20 kneadings of the separatrix of the saddle on the grid of $1000\times 1000$ points in the $(\alpha,\lambda)$-parameter region. Getting the mapping took a few hours on a high-end workstation without any parallelization efforts.
The color scan reveals a plethora of primary, large, and small scale T-points as well as the saddles separating spiral structures.

The solid-color zones in the mapping correspond to simple dynamics in the model. Such dynamics are due to either the separatrix converging to the stable equilibria or periodic orbits with the same kneading invariant
(blue region), or to the symmetric and asymmetric stable figure-8 periodic orbits (brown region). The borderlines between the simple and complex dynamics in the Shimizu-Morioka model are clearly demarcated. On the top is the curve, $LA_1$, (see the top panel of Fig.~\ref{fig5l}). The transition from the stable 8-shaped periodic orbits to the Lorenz attractor (through the boundary, $LA_{2}$) is similar though more complicated as it involves a pitch-fork bifurcation and bifurcations of double-pulsed homoclinics, see \cite{ASHIL93,SST93} for details.

One can clearly see the evident resemblance between both diagrams found using the bifurcationaly exact numerical methods and by scanning the dynamics of the model using the proposed kneading invariant technique. The latter reveals a richer structure providing finer details. The structure can be enhanced further by examining longer tails of the kneading sequences. This allows for the detection of smaller-scale spiral structures within scrolls of the primary T-vortices, as predicted by the theory. 

\section{6D optically pumped laser model}
\label{sec:6}

The coexistence of multiple T-points and accompanying fractal structures in the parameter plane is a signature for systems with the Lorenz attractor. The question remains whether the new computational technique will work for systems of dimensions higher than three. In fact, to apply the technique to a generic Lorenz-like system, only wave forms of a symmetric variable progressing in time, that consistently starts from the same initial condition near the saddle is required. Next is an example from nonlinear optics  --
a 6D model of the optically pumped, infrared red three-level molecular laser \cite{Moloney1989,FMH91} given by
\begin{equation}
\begin{array}{lll}
\dot{\beta} &=& -\sigma \beta + g p_{23}, \\
\dot{p}_{21} &=& -p_{21}-\beta p_{31}+\alpha D_{21}, \\
\dot{p}_{23} &=& -p_{23}+\beta D_{23} -\alpha p_{31}, \\
\dot{p}_{31} &=& -p_{31}+\beta p_{21} + \alpha p_{23}, \\
\dot{D}_{21} &=& -b(D_{21}-D_{21}^0) - 4 \alpha p_{21} -2 \beta p_{23}, \\
\dot{D}_{23} &=& -b(D_{23}-D_{23}^0) - 2 \alpha p_{21} -4 \beta p_{23}.
\end{array}\label{laseH}
\end{equation}
Here, $\alpha$ and $b$ are the Rabi flopping quantities representing the electric field amplitudes at pump and emission frequencies. The parameter $\alpha$ is a natural bifurcation parameter as it is easily varied experimentally. The second bifurcation parameter,  $b$, can be varied to some degree at the laboratory by the addition of a buffer gas. This system presents, like the Lorenz equations, a symmetry
$(\beta, p_{21}, p_{23}, p_{31}, D_{21}, D_{23}) \leftrightarrow  (-\beta, p_{21}, -p_{23}, -p_{31}, D_{21}, D_{23})$. The laser model has either a single central equilibrium state, $O$ (with $\beta = 0$),  or  through a pitch-fork bifurcation, a pair of symmetric equilibrium states, $O_{1,2}$ (with  $\beta \ge 0$);  the stability of the equilibria depends on the parameter values.

\begin{figure}[!htb]
 \begin{center}
\includegraphics[width=.5\textwidth]{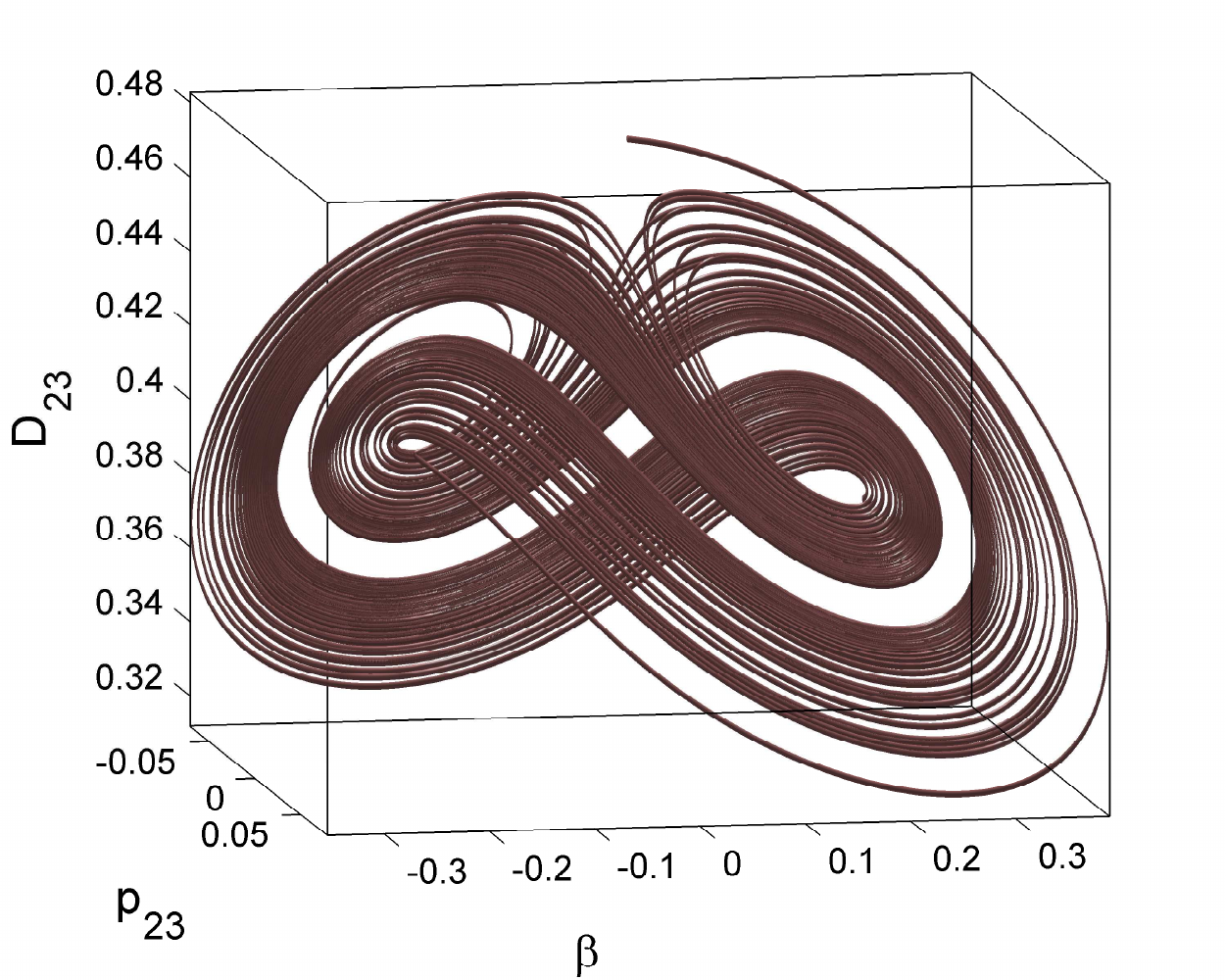}~\includegraphics[width=.45\textwidth]{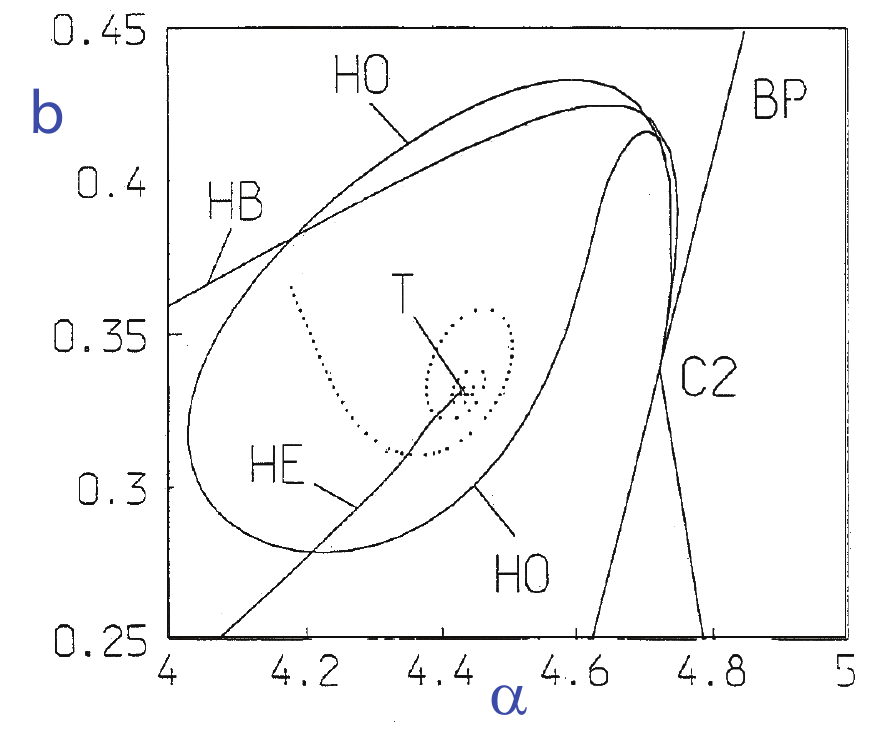}
$  $\caption{\label{laser1} (A) Lorenz attractor with a lacuna in the laser model at $a=1.14$, $b=0.2$, $q=50$ and $\sigma=10$.
 (B) $(\alpha, b)$-bifurcation diagram of the model for $g=52$ and $\sigma=1.5$. $BP$ and HB here denote the pitch-fork and Andronov-Hopf bifurcations, respectively. $HO$ and $HE$ denote the branches of the primary homoclinic (of the saddle) and heteroclinic orbits (of the saddle-foci). $C2$ is the codimension-two Khorozov-Taken point for the equilibrium state with double zero eigenvalues, and $T$ is the primary terminal point. The spiraling curve connects the T-point with the homoclinic resonant saddle on $HO$, near which separatrix loops are double pulsed ones. Courtesy of \cite{FMH91}.}
 \end{center}
\end{figure}

Optically pumped, infrared lasers are known to demonstrate a variety of nonlinear dynamic behaviors, including Lorenz-like chaos \cite{laser95}. An acclaimed
example of  the modeling studies of chaos in nonlinear optics is the two level laser Haken model \cite{Haken75} to which the Lorenz equation can be reduced. A validity that three level laser models would have the Lorenz dynamics  was widely questioned at the time. It was comprehensively demonstrated  \cite{FMH91} in 1991 that this plausible laser model possesses a wealth of dynamical and structural features of Lorenz-like systems, including the Lorenz attractor {\it per se} (including lacunae), similar Andronov-Hopf, $Z_2$ pitchfork, and various homoclinic and heteroclinic bifurcations including codimension-two   T-points, see Fig.~\ref{laser1}. Similar structures were also discovered in another nonlinear optics model for a laser with a saturable absorber which can be reduced to the Shimizu-Morioka model near a steady state solution with triple zero exponents \cite{VV93}

\begin{figure}[!htb]
 \begin{center}
\includegraphics[width=1.\textwidth]{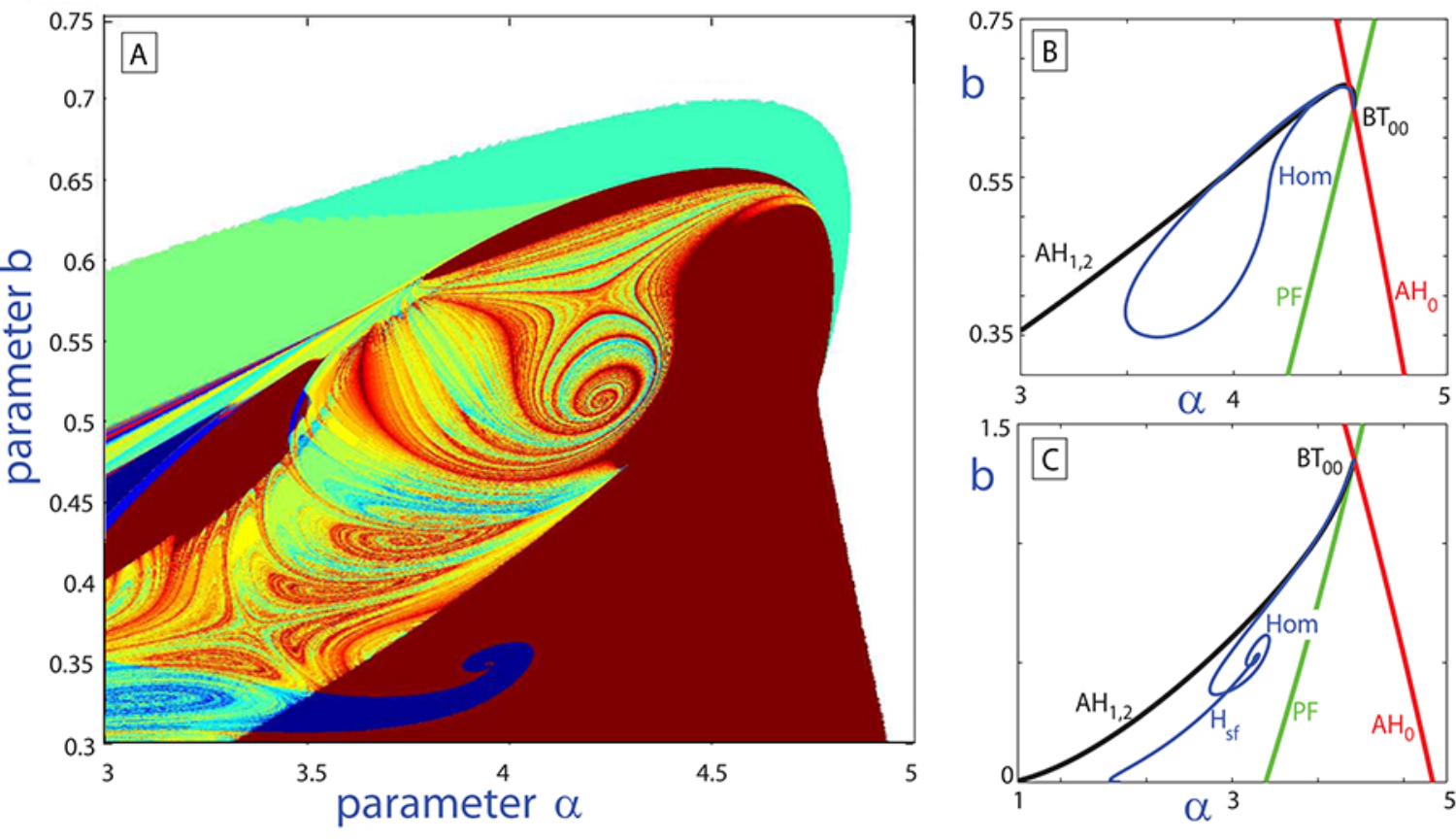}
\caption{\label{laser2} (A) Bi-parametric scan of the laser model featuring the T-points and saddles
typical for the Lorenz-like systems, mapping the dynamics of the 6D optically pumped infrared-red laser model onto the (electric-field-amplitude, omission-frequency)-diagram  at $g=50$ and $\sigma = 1.5$. Solid-color windows and fractal regions correspond to trivial and chaotic dynamics generated by the laser model. (B) Partial bifurcation diagram though the parameter continuation showing the curves for pitch-fork ($PF$) and Andronov-Hopf ($AH_0$) bifurcations for the equilibrium state, $O$, and another similar supercritical one for  $O_{1,2}$. The homoclinic curve, $Hom$ begins from the codimension-two point, $BT$ for the Khorozov-Takens bifurcation and ends up at the resonant saddle point. (B) Elevating $\sigma =2$ makes the $Hom$ turned by a saddle point in the parameter plane and terminate at the primary T-point.}
 \end{center}
\end{figure}

Likewise the Shimizu-Morioka model, the laser model~(\ref{laseH}) is rich in bifurcations. The list includes Andronov-Hopf bifurcations of equilibria, a pitch-fork  bifurcation of periodic orbits, and various curious two homoclinic bifurcations of the saddle, as well as heteroclinic  connections between the saddle and saddle-foci. Many of these bifurcations curves originate from a codimension-two Khorozov-Takens bifurcation of an equilibrium state with two zero Lyapunov exponents. 

The panel (A) in Fig.~\ref{laser2} represents the kneading scans of the dynamics of the laser model which is mapped onto the $(\alpha, b)$-parameter plane with $g=50$ and $\sigma = 1.5$. The scan is done using the same 50 kneading entries. It has regions of chaotic dynamics clearly demarcated from the solid color windows of persistent kneadings corresponding to trivial attractors such as stable equilibria and periodic orbits. The region of chaos has a vivid fractal spiral structure  featuring a chain of T-points. Observe also a thin chaotic layer bounded away from the curve $Hom$ by a curve of double-pulsed homoclinics with the kneading $\{1,-1, 0\}$ connecting the codimension-two points:  the resonant saddle and the orbit-flip both on $Hom$. One feature of these points is the occurrence of the Lorenz attractor with one or more lacunae \cite{ABS83,ASHIL93}. A strange attractor with a single lacuna containing a figure-8 periodic orbit in the phase space of the given laser model is shown in panel~A of Fig.\ref{laser1}.

\section{Conclusions}

We have demonstrated a new computational toolkit for thorough  explorations of chaotic dynamics in three exemplary models with the Lorenz attractor. The algorithmically 
simple yet powerful toolkit is based on the scanning technique that maps the dynamics of the system onto the bi-parametric plane. The core of the approach is the evaluation of the kneading invariants for regularly or chaotically varying alternating patterns of a single trajectory -- the separatrix of the saddle singularity in the system. In the theory, the approach allows two systems with structurally unstable Lorenz attractors to be conjugated with  a single number -- the kneading invariant. The kneading scans unambiguously reveal  the key features in Lorenz-like systems such as a plethora of underlying spiral structures around T-points, separating saddles in intrinsically fractal regions corresponding to complex chaotic dynamics.  We point out that no other techniques, including approaches based on the Lyapunov exponents, can reveal the discovered parametric chaos with such stunning clarity and beauty. 

The kneading based methods shall be beneficial for detailed studies of other systems admitting reasonable symbolic descriptions, including symmetric and asymmetric \cite{ALSLP91} systems of differential and difference equations  that require two and more kneading invariants for the comprehensive symbolic description.

\section{Acknowledgments}

This work is supported by the Spanish Research project MTM2009-10767 (to R.B.), and by NSF grant DMS-1009591, and MESRF project 14.740.11.0919 (to A.S). We thank Aaron Kelley, Sajiya Jalil and Justus Schwabedal for multi-hour fruitful discussions.

\end{document}